\documentclass[graybox]{svmult}

\usepackage{mathptmx}       
\usepackage{helvet}         
\usepackage{courier}        
\usepackage{type1cm}        
\usepackage{makeidx}         
\usepackage{graphicx}        
\usepackage{multicol}        
\usepackage[bottom]{footmisc}

\newcommand{\msun}{\mbox{$M_{\odot}$}}
\newcommand{\Msun}{\mbox{$M_{\odot}$}}
\newcommand{\lsun}{\mbox{$L_{\odot}$}}
\newcommand{\Lsun}{\mbox{$L_{\odot}$}}

\newcommand{\teff}{\mbox{$T_{\rm eff}$}}
\newcommand{\Teff}{\mbox{$T_{\rm eff}$}}
\newcommand{\tin}{\mbox{$T_{\rm in}$}}

\newcommand{\vinf}{\mbox{$v_{\infty}$}}

\newcommand{\mdot}{\mbox{$\dot{M}$}}

\newcommand{\msunyr}{\mbox{$M_{\odot} {\rm yr}^{-1}$}}

\newcommand{\kms}{km s$^{-1}$}

\makeindex             

\begin{document}

\title*{Eta Carinae and the Luminous Blue Variables}
\author{Jorick S. Vink}
\institute{Jorick S. Vink \at Armagh Observatory, College Hill, 
Armagh BT61 9DG, Northern Ireland, \email{jsv@arm.ac.uk}}
%
%
\maketitle

\abstract{We evaluate the place of Eta Carinae ($\eta$ Car) among the class 
of luminous blue variables (LBVs) and show that the LBV phenomenon is 
not restricted to extremely luminous objects like $\eta$ Car, but  
 extends  luminosities as low as 
log $(L/\Lsun)$ $\sim$ 5.4 -- corresponding to initial masses $\sim$25 $\Msun$, and 
final masses as low as $\sim$10-15$\Msun$. 
We present a census of S\,Doradus variability, and  discuss basic LBV properties, their 
mass-loss behavior, and whether at maximum light they form pseudo-photospheres. 
We suggest that those objects that exhibit giant $\eta$ Car-type eruptions are 
most likely related to the more common type of S\,Doradus variability.  
Alternative atmospheric models as well as sub-photospheric 
models for the instability  are  presented, but the true nature of the LBV phenomenon 
remains as yet elusive. 
We end with a discussion of the evolutionary status of LBVs -- highlighting 
recent indications that some LBVs may be in a direct pre-supernova state,   
in contradiction to the standard paradigm for massive star evolution.}

\section{Introduction}
\label{sec:intro}
 
Luminous Blue Variables (LBVs) are evolved, luminous hot stars that 
experience eruptions or outbursts and episodes of enhanced mass loss.
During outburst they appear to make transits in the HR 
Diagram (HRD) from their normal hot quiescent state to lower temperatures. 
The LBVs include a number of famous stars such as P Cyg, S Dor, R127 and AG Car. 
Eta Car is often described as an LBV, although it is a more extreme
example owing to its giant eruption.    

During the late 1970's, it was recognized that the distribution of the most luminous hot 
stars on the HRD defines a locus  
of declining luminosity with decreasing temperature \cite{Hump79,Hutch76,Sterken78}. 
Together with the fairly tight upper luminosity limit of the yellow and red supergiants at 
log ($L/\Lsun$) $\approx$ 5.8 \cite{Hump79}, this indicated that the most massive stars ($M$ $>$ 60$\msun$) do not 
evolve to cooler temperatures: the Humphreys-Davidson (HD) limit \cite{Hump79}. 
Humphreys and Davidson \cite{Hump79,Hump84} suggested that high mass-loss episodes, represented by stars 
like $\eta$ Car, P~Cyg, S~Dor and the Hubble-Sandage variables in M31 and M33 \cite{Hubble53},  
prevented the evolution of the most massive stars to cooler temperatures. 
With this addition of a post-main sequence period of high mass loss ($10^{-5} - 10^{-3}$ $\msunyr$)
the evolutionary tracks of the most massive stars were shown 
to turn bluewards, towards the core He-burning Wolf-Rayet (WR) phase.
During the WR phase, stars are anticipated to explode as supernovae (SNe) type Ib/c. However, at the end of this 
chapter, we challenge the canonical view that LBVs are always transitional between the O and WR stars, and 
suggest that some massive stars may already suffer their final explosion during or at the 
end of the LBV phase. One of the most relevant questions is therefore {\it do LBVs explode?} 

The term LBV was introduced to describe the diverse group of unstable 
evolved hot stars in the upper HRD. Today we distinguish between more than one type of 
LBV \cite{Hump94}: (i) the normal LBV variability cycles with visual magnitudes 
changes of 1-2 magnitudes at essentially constant luminosity -- on timescales of years to decades -- 
represented by the prototype of the class, S~Dor, in the Large Magellanic Cloud, 
 and (ii) the giant eruption LBVs represented by $\eta$ Car and P~Cyg with visual magnitudes 
changes of 3 magnitudes or more, during which the total bolometric luminosity 
increases \cite{Hump94,Hump99}.   

In this chapter, we focus on the S~Dor-type variables and their transits in the HRD, the prime 
characteristics of the LBV class. Roughly 30 massive stars in the Galaxy ($\simeq$10) and  
Local Group ($\simeq$20) are known to be S\,Dor variables.  
By contrast, only two Galactic objects have been discovered to exhibit giant $\eta$ Car-like 
eruptions: $\eta$ Car itself and P~Cygni, which
suddenly appeared at naked-eye visibility in 1600. Due to their high luminosities at 
maximum, a significant number of LBV-like non-terminal eruptions have   
been discovered in external galaxies. Most are typical of giant eruptions whilst 
some appear to be more similar to S\,Dor type variables. 
These extragalactic LBVs are sometimes referred to as ``SN imposters'' 
(see the contribution by Van Dyk and Matheson, this volume).
 
The circumstellar nebulae seen around many Galactic and 
MC LBVs \cite{Nota95,Weis03} may also have resulted from 
giant $\eta$ Car eruptions, although stationary winds from prior evolutionary 
phases may constitute an alternative scenario for their creation \cite{Garcia96}. Given the association 
of many LBVs with nebulae and ejecta and the close proximity of both the S Dor variables and 
the $\eta$ Car-like variables to the 
Eddington limit, it is often suggested (though not proven) that their instabilities 
represent different manifestations of the same underlying evolutionary state. The 
fact that the giant eruptor P~Cyg is also subject to 
small amplitude S\,Dor variability \cite{deGroot01,Markova01} 
and  that $\eta$ Car's second outburst (during 1888-1895) was like that of a 
normal S\,Dor variable \cite{Hump99} lends support to this possibility. 

We emphasize, however, that neither the $\eta$ Car-type eruptions, nor the S\,Dor variations 
are understood. Worse still, we do not know whether all LBVs are subject to both 
types of variability, or in which order they may occur.  
Two pertinent questions are thus whether ``{\it $\eta$ Car is 
unique among the LBVs?}'' and ``{\it {what is the root cause of the S~Doradus 
variations?}}''


\section{Basic properties of LBVs}
\label{sec:props}

\noindent {\bf Variability}~~ LBVs show significant spectroscopic and photometric 
variability  on timescales of years (short S\,Dor phases) to decades 
(long S\,Dor phases, cf. van Genderen \cite{vanG01}).
For completeness, we note that LBVs also show smaller amplitude
``micro'' variability, on shorter timescales (weeks, months), but this aspect
of their variability is  also a common feature of supergiants in general
(see \cite{Lucy76,Lef07} and references therein).\\ 

\begin{figure}
\sidecaption
\includegraphics[scale=.65]{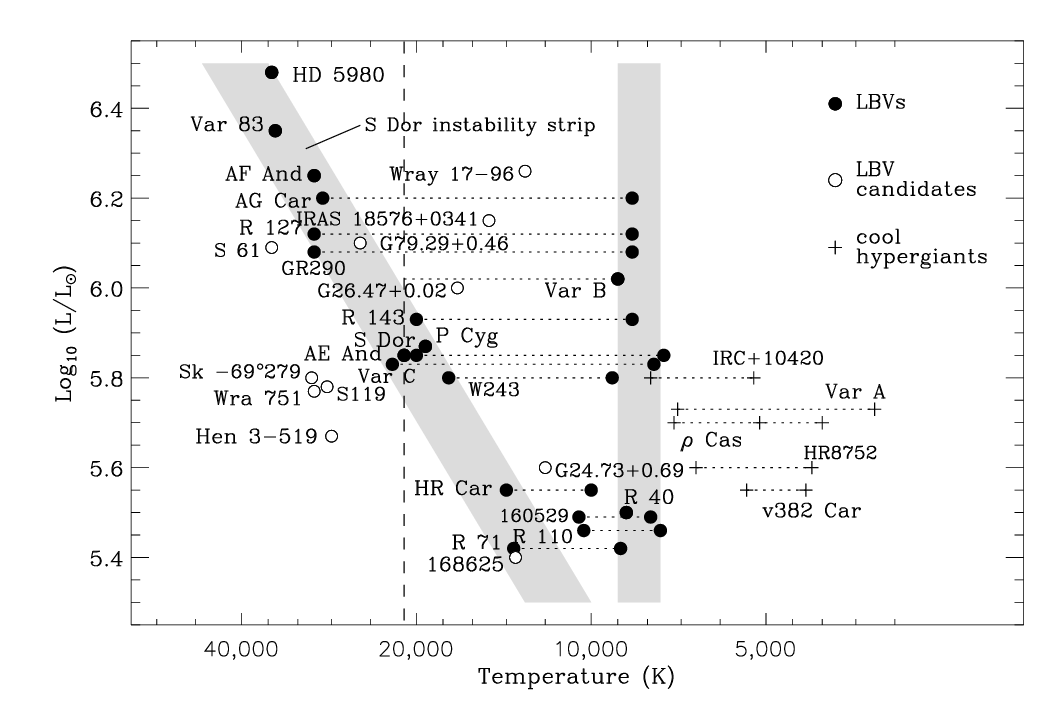}
\caption{The location of the LBVs (black circles) and candidates (open circles) in 
the Hertzsprung-Russell Diagram. The cool yellow hypergiants are indicated with 
pluses. The slanted and vertical grey bands represent visual minimum and maximum 
respectively. The dashed vertical line at 21\,000K indicates the location of the 
bi-stability jump (see Sect.~\ref{sec:mdot}). The figure has been adapted from 
Smith et al. \cite{Smith04} and is similar to e.g. Fig.~9 in \cite{Hump94}.}
\label{fig:hrd}
\end{figure}

\noindent {\bf Luminosities}~~ 
The classical LBVs have log ($L/\lsun$) larger than 5.8 with bolometric magnitudes in the range $M_{\rm bol}$ $-$9 to $-$11.
There is an apparent ``gap'' in their luminosities just below log ($L/\lsun$) $=$ 5.8,
and a separate group of less luminous LBVs with log ($L/\lsun$) $=$ 5.4-5.6, corresponding
to bolometric magnitudes in the range $M_{\rm bol}$ $\simeq$$-$8 to $\simeq$$-$9 \cite{Hump94}. 
We note that this separation in luminosity (see Fig.\ref{fig:hrd}) may 
not be real, but due to small number statistics.
As the less luminous LBVs are below the HD limit, they have presumably been red supergiants where they may already have 
shed a lot of mass. As a result, they may be equally close to their Eddington limit as the classical LBVs.

One of the most important properties of LBVs is that they appear to be close to 
the Eddington limit for stability against radiation pressure for their luminosities and current
masses (see the subsection on stellar masses). The Eddington luminosity $L_{\mathrm Edd}$ is 
defined:    

   \begin{equation}  
     L_{\mathrm{Edd}} \ = \ \frac{4{\pi}cGM}{{\kappa}_{\mathrm{F}}}    
   \end{equation}  
where ${\kappa}_{\mathrm{F}}$ is the flux-mean opacity and $M$ is 
the total mass of the star. 

The dimensionless Eddington parameter $\Gamma$ is defined as the 
ratio of radiative acceleration to gravitational acceleration 
(disregarding sign):

  \begin{equation}  
  {\Gamma}  \  =  \   
      \frac{ {\kappa}_{\mathrm{F}} L }{ 4 {\pi} c G M }   \ = \ 
        (7.66 \times 10^{-5} \; \mathrm{g} \; \mathrm{cm}^{-2}) 
            \, {\kappa}_{\mathrm{F}} \; \left( \frac{L}{L_{\odot}} \right)
            \; \left( \frac{M}{M_{\odot}} \right)^{-1} \ .     
  \end{equation}  
If opacity is entirely due to Thomson scattering by free electrons, 
as often assumed in this connection, then 
     ${\kappa}_{\mathrm{F}} \; \approx \; 0.3$ cm$^{2}$ g$^{-1}$ ;    
but this depends somewhat on chemical composition and ionization 
state.

\noindent {\bf Temperatures}~~ LBV temperatures are not only  
 time-variable, but accurate $\teff$ determination requires 
sophisticated non-LTE spectral analyses for hot stars with extended envelopes and winds. 
Furthermore, it should be noted that for extreme objects like $\eta$ Car we cannot even estimate the radius based on 
$\Teff$ as the wind is optically thick. In such a case, $\Teff$ refers to the position in the wind where $R_{\rm ph}$ $>$ $R_{\rm sonic}$
(see below for a more extended discussion). 

Fig.~\ref{fig:hrd} shows the confirmed and candidate LBVs on a schematic HRD with 
their transits (the dotted horizontal lines) between quiescence 
and their outburst state or visual maximum.   
At quiescence, or visual minimum, they appear to lie along a fairly narrow slanted 
band: the ``S Dor instability strip'' \cite{Wolf89}. Their effective temperatures vary 
from over 30~000 K, corresponding to spectral type late O/early B, for the luminous classical 
LBVs to only $\simeq$15~000 K, corresponding 
to late B spectral types, for the less luminous LBVs. Objects from both subgroups 
transit redwards to 
$\teff$ not higher than $\simeq$8~000 K, corresponding to spectral types  A to F, 
when they are ``in outburst'' or visual maximum.  
Wolf \cite{Wolf89}  noted that the amplitudes of the S~Dor excursions become larger with 
increasing luminosity, and suggested that the  
most luminous LBVs could be cosmological distance indicators. 
Normal blue supergiants are also observed in and to the right of 
the S\,Dor instability strip, but they may not be close to their Eddington limit and may 
therefore not be subjected to the instability that characterizes 
LBVs. The cool or yellow hypergiants (YHGs)  just below the HD limit 
\cite{Hump79, Hump94} are included as these might somehow represent the ``missing'' LBVs \cite{Smith04} -- a 
possibility which will be discussed further in Sect.~\ref{sec:mdot}.\\

\noindent {\bf Membership} Whether a particular blue hypergiant is a member of 
the select group of LBVs (see Table~\ref{tab:lbv}) is often a matter of  
debate (see \cite{Hump94, Clark05, Massey07}). In general, spectral resemblance to the 
P\,Cygni type spectrum of a known LBV is not a sufficient  criteria for a  
star to be called an 
LBV. For this reason we include a group of LBV candidates (LBVc) listed in 
Table~\ref{tab:lbvc}. To officially qualify as an LBV, an object should at 
least have shown the combination of spectral-type/photometric variations characteristic 
of S~Dor-type variability due to changes in the star's {\it apparent} temperature. Spectral variability, 
e.g. due to mass-loss changes, on its own is not a criterion, as almost all 
massive stars would qualify. Changes in photometric color  could  be the 
result of $\teff$ changes, however obscuration by circumstellar dust may also 
give rise to color changes. 
Finally, the association of a supergiant with a circumstellar 
nebula is also not a criterion for an object to be added to the list of 
LBVs. The famous pistol star and LBV 1806$-$20 are thus strictly not 
LBVs. These objects remain candidates until they have shown the characteristic spectral-type 
variations.

\begin{table}
\caption{The confirmed LBVs. See \cite{Valeev09,Mauerhan10,Maryeva11} for the latest additions.}
\label{tab:lbv}       
\begin{tabular}{p{2cm}p{2cm}p{2cm}p{2cm}p{2cm}}
\hline\noalign{\smallskip}
{\it Galaxy:~~}\\ 
$\eta$ Car & AG Car    & HR Car  & P Cygni     & HD 160529\\ 
HD 168607  & FMM 362 & AFGl 2298 & G24.73$+$0.69 &  W1-243\\       
GCIRS 34W  & Wra 751 & G0.120 - 0.048\\
{\it LMC:~~}\\
S Dor      & R71       & R 110    & R 116    & R127\\    
R 143      & HD 269582 & HD 269929\\
{\it SMC:~~}\\
R40        & HD 5980\\
{\it M31:~~}\\
AE And    & AF And   & Var A-1 & Var 15\\
{\it M33:~~}\\
Var B     & Var C    & Var 2   & Var 83 & GR\,290 \\ 
N\,93351\\
{\it M81:~~}\\
I 1   & I 2          & I 3 \\
{\it M101:~~}\\
V 1   & V 2          & V 10\\
{\it NGC 2403:~~}\\
V 12    & V 22       & V 35    & V 37   & V38\\
{\it NGC 1058:~~}\\
SN 1961 V\\
\noalign{\smallskip}\hline\noalign{\smallskip}
\end{tabular}
\end{table}

\begin{table}
\caption{The LBV candidates. See \cite{Sterken08,Gvaramadze10,Buemi11} for the latest additions}
\label{tab:lbvc}       
\begin{tabular}{p{2cm}p{2cm}p{2cm}p{2cm}p{2cm}}
\hline\noalign{\smallskip}
{\it Galaxy:~~}\\ 
Cyg OB2\#12 & Pistol star  & HD 168625     & HD 326823     & HD 316285\\
He3-519     & HD 80077     & $\zeta^1$Sco  & MWC 314       & MWC 930\\
AS 314      & G25.5$+$0.2  & G79.29$+$0.46 & G26.47$+$0.02 & Wra17-96\\
WR102ka     & LBV1806$-$20 & Sher 25       & W51 LS1       & GCIRS 16NE\\  
GCIRS 16C   &GCIRS 16SW    & GCIRS 16NW    & GCIRS 33SE \\ 
MN112 \\
{\it LMC:~~}\\
R 4         & R 66         & R 74          & R 78         & R 81\\
R 84        & R 85         & R 99          & R 123        & R 128\\
R 149       & S 18         & S 22          & S 61         & S 119\\
S 134\\
\noalign{\smallskip}\hline\noalign{\smallskip}
\end{tabular}
\end{table}

When LBVs are discussed in an evolutionary context, we should be 
aware that the LBV phenomenon is likely to be intermittent and that part of the 
population might be dormant. Disregarding this could lead to 
incorrect interpretations with respect to their relative numbers and their evolutionary state. 
When we discuss the LBV phenomenon, however, 
we should only include the confirmed LBVs.\\  

\noindent {\bf Abundances}~~As massive stars evolve on the main sequence their atmospheric 
abundances are expected to   undergo 
a transition in chemical abundances from solar He and CNO abundances to He-enriched 
and nuclear-equilibrium CNO abundances (with N enhanced, 
C/O depleted) \cite{Maeder83}. For  massive stars with $\simeq$60 $\msun$, this transition occurs 
rather rapidly after about 2 Myr of evolution \cite{Yungel08}. 
LBVs show a wide span of CNO ratios.  Davidson et al. \cite{Dav82} showed that the 
ejecta around  $\eta$ Car are enhanced in both nitrogen and helium, and  Pasquali 
et al. \cite{Pasquali02} found that the shell ejected by the LBVc HD\,168625 is also N-enriched compared to the interstellar medium by a factor 
of several. Smith et al. \cite{Smith98} studied a sample of LBV nebulae and found the ejecta to be generally N-enhanced, although most 
nebulae have not reached abundances characteristic of CNO equilibrium. 

The origin of the circumstellar nebulae associated with some LBVs however is uncertain. 
Photospheric abundance 
measurements might therefore provide a more direct means to constrain the LBV evolutionary state. These 
abundances however depend not only on the details of the atomic physics but also on the  
complexities of non-LTE spectral analyses.  
For example,  Hillier et al. \cite{Hillier01} showed that the H/He ratio 
in the wind of $\eta$ Car is ambiguous due to a strong coupling with the mass-loss rate. 
Eta\,Car however, could be an exception  with its huge luminosity and high  mass-loss 
rate.  Najarro et 
al. \cite{Najarro97} studied the atmospheric He abundance of P\,Cygni and quoted a best value of
n(He)/n(H) of 0.3, corresponding to a mass fraction Y $=$ 0.63. However, they noted 
 a huge uncertainty due to the trade-off in ionization and He abundance and provided  
a He abundance range of n(He)/n(H) $=$ 0.25-0.55. Hillier et al. \cite{Hillier98} also ruled out 
a solar He abundance for the extreme P~Cygni star HD\,361285, but admitted that the 
uncertainty in n(He)/n(H) could  be as large as a factor 20! 
Evidence for advanced CNO processing was found by Lennon et al. \cite{Lennon93} for the 
LBV R71, but this is not a well-established result for LBVs in general.  
Although most LBVs have He and N enhanced atmospheres, it seems unlikely that  
all of them have reached equilibrium CNO values in their outer atmospheres.\\

\noindent {\bf Stellar masses}~~Masses are most accurately determined using 
detached binary systems. However, almost all LBVs are single. Eta Car is the 
best-known exception to this rule, with its 5.5y periodicity \cite{Whitelock94,Dami96} 
attributed to  a companion. 
The current  mass of $\eta$\,Car is  thought to be at least 
90 $\Msun$, to avoid violating the Eddington Limit at its high luminosity (see Davidson this volume, Owocki and
Shaviv this volume). 

 LBVs with log ($L/\Lsun$) $>$ 5.8, are usually assumed to be evolved from  
 the most massive stars with 
$M$ $>$ 50$\msun$ (e.g. \cite{Schaller92}). Mass measurements for these 
high-luminosity objects are scarce and uncertain. Pauldrach \& Puls \cite{Pauldrach90} quote a 
mass of 23 $\msun$ for P~Cyg. Vink \& de Koter \cite{Vink02}  estimated  
LBV masses from time-variable mass-loss rates and found a mass of 35 
$\msun$ for AG Car. These results  suggest that classical LBVs  have 
already lost a significant fraction of their initial mass probably through the combined effects 
of line-driven mass loss during the OB and LBV phase, and/or through 
 prior major eruptions.
The quoted mass estimates are highly model dependent. For instance,  
wind clumping was not considered for the mass estimate of AG\,Car. 
However, if these objects have lost half their initial mass, their $L/M$ ratio 
is  quite large, with $\Gamma_e$ $\simeq$ 0.5, and they are  close to the 
Eddington Limit for their luminosities.  

The initial masses of the less luminous LBVs,  log ($L/\Lsun$) $\simeq$ 5.4,  is of 
the order $M$ $\sim$ 25$\msun$; e.g. \cite{Schaller92}). Comparing this mass with 
current-day mass estimates of $M$ $\sim$ 12$\msun$ for R71 \cite{Leit89}, $M$ 
$\sim$ 10$\msun$ for R110 \cite{Stahl90}, and $M$ $\sim$ 13$\msun$ for HD\,160529 
\cite{Sterken91}, suggests that these low-luminosity LBVs may already have 
lost more that half of their initial mass (e.g. during a prior RSG phase). 
The low-luminosity LBVs  are  equally close to their  Eddington limit 
 with $\Gamma_e$ $\simeq$0.6. All of these  empirical mass 
estimates are  uncertain by at least a factor two.

Interestingly, Martins et al. \cite{Martins06} recently reported that the LBVc GCIRS 16SW  may be  
an eclipsing binary, with both components weighing $\simeq$50 $\msun$. If it is  
 a genuine LBV, it may  be a key object for constraining massive star evolution models, 
 because current models with rotation \cite{Meynet05} suggest such massive objects will not pass 
through the LBV phase at all during their evolution.


\section{Mass-loss properties -- do LBVs form pseudo-photospheres?}
\label{sec:mdot}

With its current mass-loss rate of order 10$^{-3}$ $\msunyr$ 
 it is clear that $\eta$\,Car has formed an optically-thick wind and a pseudo-photosphere, but whether 
the S\,Dor-type variables 
have optically thick winds is less well established. 
It has been suggested that the temperature changes during S~Dor cycles do not represent 
true stellar temperature changes, but are due to the formation of the dense, optically-thick wind. 
As the mass-loss rate increases, the effective photosphere of the moves out into the wind,  
and the apparent effective temperature of the star drops, whilst the apparent stellar 
radius increases -- without an actual expansion of the star \cite{Ap86,Lamers86,Dav87}. Davidson \cite{Dav87} also 
showed that the minimum temperature the wind can achieve
as the mass loss increases is $\sim$ 7000K, in rough agreement with the apparent temperatures of the 
LBVs at visual maximum.

Normal OB stars have winds that are optically thin in the continuum and 
we see through the entire wind -- down to the photosphere. In other 
words, the wind is formed outside the photosphere. However, when the mass-loss rates increases, the 
wind becomes less transparent, and the photosphere from which the optical light originates  is now at larger radii.
When the wind has become optically thick, the photosphere is formed above the sonic velocity. 
In other words, the wind is accelerated ``inside'' the photosphere -- forming an opaque wind with 
$R_{\rm ph}$ $>$ $R_{\rm sonic}$.

Leitherer et al. \cite{Leit89} and de Koter et al. \cite{deKoter96} performed 
detailed NLTE modelling of this process, showing that the extent of a pseudo-photosphere 
that results from increased mass loss is relatively modest in comparison to that 
of a WR star.  The conclusion was that the underlying LBV radius itself must become 
larger due to an as yet unidentified sub-photospheric mechanism. Furthermore, it became
clear that empirical mass-loss rate during S\,Dor redward excursion do not always 
increase during outburst \cite{Leit97,Vink02}. 
More recently, Smith et al. \cite{Smith04} showed that pseudo-photospheres 
might be feasible under certain special circumstances
discussed later in Sect.\ref{s_pseudo}. Before we provide 
a more detailed account of the possibility of pseudo-photosphere 
formation, we give an overview of our current knowledge of stationary LBV mass loss. 
The radiative forces that may be responsible for giant eruptions are 
discussed elsewhere (Owocki and Shaviv, this volume).

\subsection{Observed mass-loss rates}

Analysis of the blue-shifted absorption components in the P-Cygni profiles of strong 
emission lines such as H$\alpha$ is commonly used to determine mass-loss rates from massive stars.\\  

\begin{figure}
\sidecaption
\includegraphics[scale=.45]{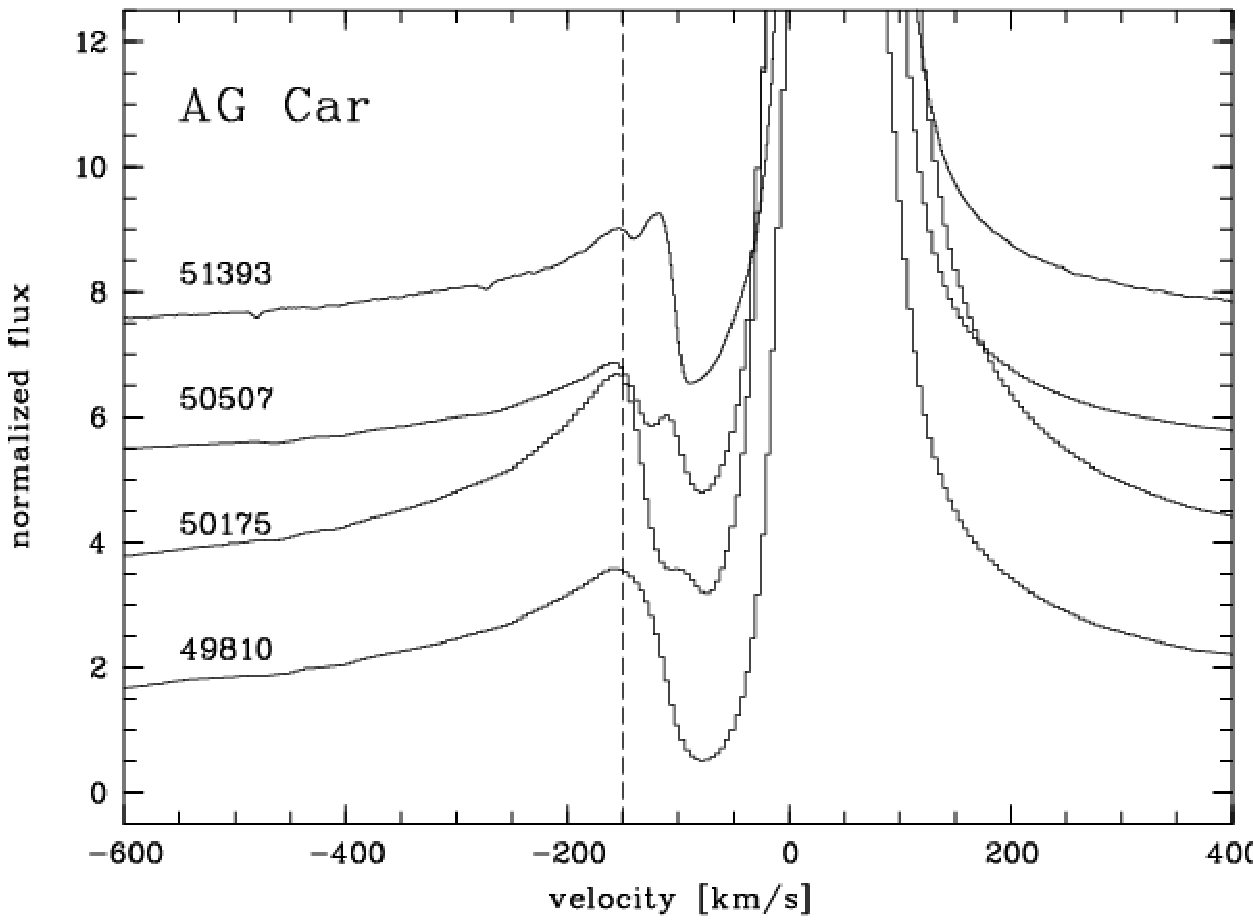}
\caption{The appearance of double-split absorption components in H$\alpha$ in the spectrum of AG~Car 
during the years 1996 to 1999. The spectra are marked in Julian days.
The figure has been taken from Stahl et al. \cite{Stahl01}.} 
\label{fig:stahl11}
\end{figure}

\noindent {\bf Terminal velocities~~} The terminal velocities of LBV winds  measured 
from the blue edge of the P~Cygni  absorption component  
are  in the range 100-250 kms$^{-1}$, with $\eta$ Car having a $\vinf$ of $\sim$500 kms$^{-1}$ 
(cf. \cite{Leit97}). These wind velocities are 
significantly lower than those of normal OB supergiants, which have $\vinf$ $\sim$1000-3000 \kms. 
The mass-loss rates of LBVs are also a factor of 10-100 larger than 
those of normal supergiants, so their wind densities, 
$\rho (r) = \dot{M}/4 \pi r^2 v(r)$, are much higher, giving the line 
profiles their characteristic P~Cygni shapes. As the  
LBV  mass-loss rate  is variable, some LBVs exhibit
profile shape changes and variability 
in the absorption  profile, 
such as the split blue-shifted absorption components seen in the H$\alpha$ line 
of AG\,Car (see Fig.~\ref{fig:stahl11}) and other LBVs 
such as R 127 \cite{Stahl83,Stahl86}.  P\,Cyg, R\,66, and HD\,160529 also exhibit 
shell components in their Fe {\sc ii} lines 
(cf. \cite{Lamers85} for P\,Cyg). The split H$\alpha$ absorption components in LBVs have recently
been proposed to be the result of an abrupt bi-stability jump \cite{groh11}.\\

\noindent {\bf Mass-loss rates~~} Mass-loss rates of most  LBVs have been determined 
using non-LTE models such as {\sc cmfgen} \cite{HillierMiller98}. The very high mass-loss rate 
for  $\eta$\,Car ($3\times$10$^{-3}$ $\msunyr$) has been determined by  several methods using radio, mm-wavelength, and Hubble Space Telescope data 
\cite{White94,Cox95,Dav95}, and the  non-LTE model results for $\eta$\,Car \cite{Hillier01}  yield a similar answer.  
Sophisticated NLTE mass-loss determinations for other LBVs include a value of 
$3\times$10$^{-5}$ $\msunyr$ for the extreme twin of 
P\,Cyg: HD\,316285 \cite{Hillier98}, while 
Najarro et al. \cite{Najarro97} derived $3\times$10$^{-5}$ $\msunyr$ for P\,Cyg itself, using similar analysis tools. 
These results are all much higher than for normal OB supergiants of comparable temperatures. 

\begin{figure}
\sidecaption
\includegraphics[scale=.65]{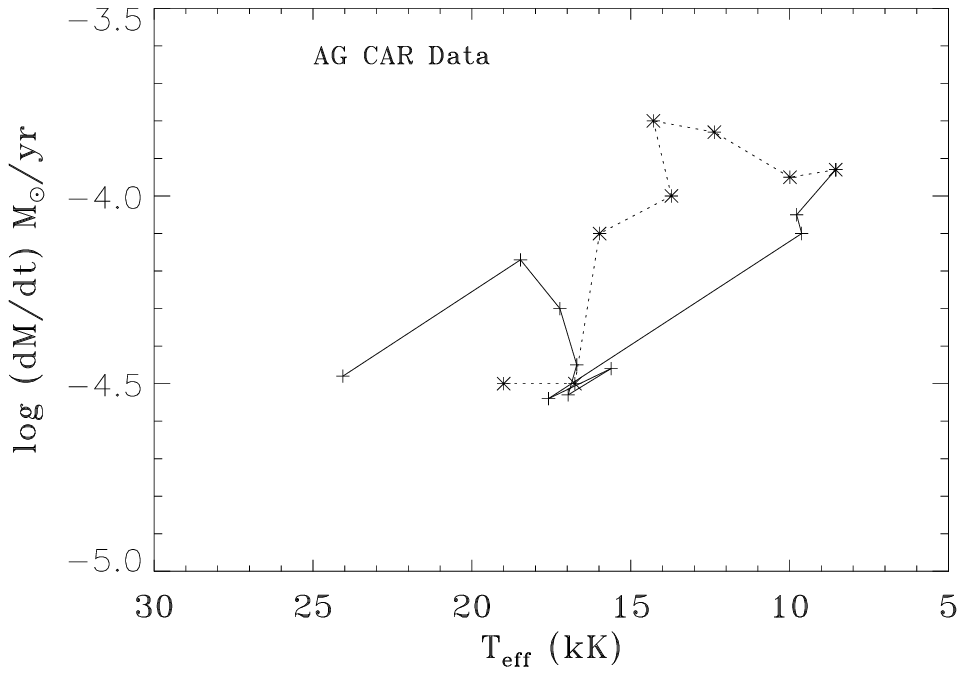}
\caption{Time-variable empirical mass-loss rates for AG~Car as a function
of $\teff$, as analyzed by Stahl et al. \cite{Stahl01}. Figure taken from Vink \& 
de Koter \cite{Vink02}.}
\label{fig:agcar}
\end{figure}

The non-LTE model based mass-loss determinations however are based on assumed spherical winds, but  the
wind of  $\eta$ Car,  is bipolar \cite{Smith03,vanBoekel03}. Whether non-spherical winds have  a large effect 
on the mass-loss determination, remains to be shown. A more serious assumption may 
be homogeneity. Davies et al. \cite{Davies05} performed a linear spectropolarimetry survey 
of Galactic and MC LBVs and found  large line polarizations in over half of their survey
targets.  This is a higher incidence of 
polarization line effects than  in O and WR stars where asphericities of resp. 
$\sim$25\% \cite{Harries02,Vink09} and $\sim$15\% \cite{Harries98,Vink07} have been reported. 
Rather than attributing the polarization to large-scale 
axi-symmetry (e.g. \cite{Schulte94}), Davies et al. \cite{Davies05} attribute the 
linear polarization of LBVs like AG\,Car to wind clumping, because the position angle in the 
polarization was shown to vary significantly, as was the case for  
P\,Cyg \cite{Nordsieck01}. A similar conclusion regarding 
overall sphericity of LBV winds was drawn by Guo \& Li \cite{Guo07} on 
the basis of modelling LBV continuum energy distributions. 

The LBV polarization variability implies that the clumps  
 must arise close to the 
photosphere, and  many small clumps predominate over a few larger 
ones \cite{Davies07a}. The quantitative 
implications of wind clumping for the absolute mass-loss 
rate of LBVs  and in  early-type stars, in general, 
have yet to be established \cite{Hamann08}. 

Although most LBVs have been monitored photometrically, only a handful have been 
subject to quantitative spectroscopic analysis at various epochs, and mass-loss 
rates have rarely been reported for different S\,Dor phases. In this respect, 
AG\,Car is  the best studied LBV. Stahl et al. \cite{Stahl01} investigated AG\,Car's 
mass-loss behavior over the period December 1990 -- August 1999 and modelled the H$\alpha$ profiles in detail. 
Their empirical mass-loss rates for the cycle from visual minimum to maximum -- and back to minimum --  
are plotted against apparent temperature in Fig.~\ref{fig:agcar}. 
The mass-loss rate rises, drops, and rises again
towards visual maximum (solid line)  due to ionization 
changes of the Fe lines that drive the wind \cite{Vink02}. 
We note that there is a difference in 
mass-loss behavior  from visual minimum to maximum and  in the 
opposite direction (dotted line). We suspect that this is due to the breakdown of the 
assumption of stationarity from outburst to quiescence. Due to the larger radii, the dynamical flow times are 
much longer at maximum than they are at minimum light, which implies that material that was 
lost in this 
phase may still be near the photosphere, which may 
significantly affect the mass-loss determinations, resulting in erroneously large 
mass-loss rates for the route back to minimum. 
A second reason for the difference may be related to the release of gravitational energy 
when the star returns to quiescence. If this plays a role, the assumption of constant 
bolometric luminosity may no longer hold (see also \cite{Groh09a,Clark09}. 
Due to the above-mentioned complexities, we focus the 
comparison of mass-loss predictions to empirical mass-loss rates for the outburst phase (solid line) only.

\subsection{Theoretical mass-loss rates}

Mass loss from a star with a stationary stellar wind is assumed to be due to  
 an outward acceleration  larger than the inward directed 
gravitational acceleration. For early-type stars, this acceleration has been 
identified with the radiation force, which depends on both the available 
photospheric flux and the cross section of the particles that can 
intercept this radiation. 

In hot-star winds, nearly all H is ionized by the strong radiation 
field, which implies that there is an enormous number of free electrons present 
which are the main contributors to the 
continuum opacity. The radiative acceleration due to photon scattering off free 
electrons is subject to the same $1/r^2$ radius dependence as is the gravitational 
acceleration, and for this very reason cannot drive a stellar wind by itself. 
Lucy \& Solomon \cite{LS70} showed that a stationary wind would occur  when  
scattering by optically thick spectral lines was included.   
The interested reader is referred to the introductory book on stellar winds by 
Lamers \& Cassinelli \cite{Lamers99}
for an overview of the line acceleration of optically thin and thick lines. 

The line acceleration due to {\it all} spectral lines is often expressed in terms 
of the radiative acceleration due to electron scattering times a certain 
multiplication factor: the force multiplier $M(t)$. Using this method, one 
can parametrize the line force, and solve the equation of motion in a rather 
straightforward manner \cite{CAK}. In this approach, the radiation is 
assumed to emerge directly from the star. The  effects of diffuse 
radiation and multiple scatterings are not taken into account. 

Abbott \& Lucy \cite{AL85} showed that calculated  mass-loss rates can also be obtained 
using Monte Carlo simulations, counting the cumulative radiative 
accelerations due to photon interactions with gas particles of different chemical 
species (mostly Fe). However, the main challenge in radiation-driven wind dynamics 
is that the line acceleration $g_{\rm line}$ depends on the velocity gradient 
$(dV/dr)$, but the velocity $V(r)$, hence $dV/dr$, in turn depends on 
$g_{\rm line}$. Due to its non-linear character, the dynamics of line-driven winds 
is quite complex. Fortunately, observational analyses provide accurate 
information on wind velocities, which can be used to constrain the wind 
dynamics. 

Vink \& de Koter \cite{Vink02} adopted an empirical velocity stratification\footnote{Note that 
our treatment of the winds dynamics has now been updated \cite{Mueller08}.}, $V(r)$, and 
predicted stationary mass-loss rates for LBVs as a function of $\teff$ in a similar 
vein to their mass-loss prescriptions for OB supergiants \cite{Vink00}. 
They studied the effects 
of lower masses and modified He/H and CNO abundances in comparison to normal OB supergiants and  
found that the main difference in mass-loss rate is attributable to the lower masses 
of LBVs compared to OB supergiants, resulting in a larger Eddington parameter 
$\Gamma_{\rm e}$.  
The increase in He abundance changed the mass-loss properties by only very 
small amounts (up to about 0.2 dex in log $\dot{M}$). CNO processing also had only 
a minor effect on the mass-loss rate, because Fe was found to be the 
dominant contributor to the line force in the inner wind. CNO lines 
contribute mostly to the line force in the outer wind where the terminal velocity 
is set \cite{Vink99,Puls00}.

\begin{figure}
\sidecaption
\includegraphics[scale=.65]{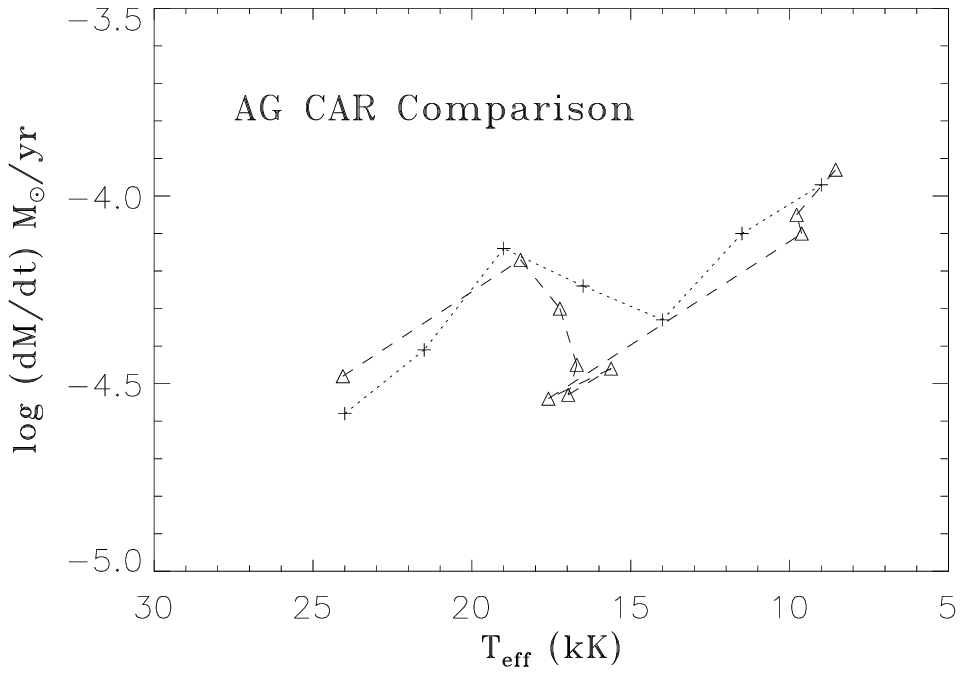}
\caption{Predicted (dotted line) and empirical (dashed line) mass-loss rates 
versus \teff\ for the LBV AG~Car. Note that both the qualitative behaviour and the 
amplitude of the mass-loss variations are well reproduced, provided the predictions 
are shifted by $\Delta \teff = -6\,000$ K. See \cite{Vink02} for details.}
\label{fig:agcomp}
\end{figure}

They \cite{Vink02} also compared their LBV mass-loss predictions with  
observational analyses and showed that the mass-loss variability during the S\,Dor 
cycles may arise from changes in the ionization balance of Fe: the bi-stability mechanism,
 first noticed in model 
calculations of the wind of P~Cygni \cite{Pauldrach90}. 
The wind either had a high $\vinf$ and low 
$\dot{M}$, or visa versa. The location of the jump near spectral type B1 (21\,000 K) 
was established by Lamers et al. \cite{Lamers95} from a $\vinf$ study of OB supergiants and 
improved by Crowther et al. \cite{Crow06}.
The nature of the jump was originally attributed to the optical thickness of the 
Lyman continuum \cite{LP91}, but Vink et al. \cite{Vink99} showed that most of the line driving for 
both the hot and cool side of the jump was due to Fe  in the Balmer 
continuum, with the jump in mass loss being the result of the recombination of Fe {\sc iv}, as 
Fe {\sc iii} has more lines available to drive the wind. 
 The first empirical evidence for a jump in $\dot{M}$ may 
have been found by Benaglia et al. \cite{Benaglia07}, even though the rates at later spectral 
types appear to drop below those predicted \cite{Vink00,Trundle05,MP08}.

The intriguing case of AG\,Car is depicted in Fig.~\ref{fig:agcomp}, where we 
compare the 
predictions with the Stahl et al. \cite{Stahl01} rates when 
the apparent temperature  decreased from 24\,000 to 9\,000 K. In these 
computations, it was assumed that log ($L/\lsun$) $=$ 6.0; $M$ $=$ 35 $\msun$; the 
He mass fraction Y $=$ 0.60, and the ratio of the terminal over escape velocity 
was 1.3. The luminosity and He abundance are similar to those assumed by Stahl 
et al. We used the mass-loss behavior to constrain the stellar 
mass of the LBV.  Unfortunately, the terminal velocity is poorly constrained by observations, 
as the $\vinf$ determination from H$\alpha$ only allowed for a lower limit \cite{Stahl01}. 
We note that the adopted ratio of the terminal over the effective 
escape velocity may impact the mass determination. Furthermore, the Fe recombination 
temperatures show an offset compared to empirical 
constraints from the drop in terminal velocities at spectral type B1 
in OB supergiants. 

Figure~\ref{fig:agcomp} shows that after accounting for a corrective shift $\Delta$ 
$\teff$, the observed and predicted mass loss agree within $\simeq$0.1 dex. As 
\mdot(\teff) shows a complex behavior, with fluctuations of over 0.5 dex, this  
is a satisfactory result, confirming that AG\,Car's 
mass-loss variability is the result of changes in the ionization of the dominant 
line-driving element Fe.

\subsection{Do S~Dor variables form pseudo-photospheres?}
\label{s_pseudo}

Now that we have gathered information on the empirical and theoretical 
mass-loss rates of LBVs, we can start addressing the question of whether these 
rates are large enough to be capable of forming a pseudo-photosphere.
In most modern 
non-LTE atmosphere codes, the core radius follows from the relation 
$L = 4 \pi R_{\rm in}^2 \sigma \tin^4$. As the inner boundary is chosen to be deep 
in the stellar photosphere the input temperature does not necessarily equal the 
output effective temperature. The effective temperature \teff\ is defined at 
the position where the thermalization optical depth at 5555 \AA\ equals 
$1/\sqrt{3}$. We intentionally choose the thermalization optical depth over 
purely thermal optical depth, as we wish to include the effects of dilution by scattering (see 
\cite{deKoter96} and references therein for more extensive discussions). 
For stars with modest mass fluxes, such as normal O stars, the winds are optically thin and \teff\ is only
slightly lower than \tin. For LBVs, with 
$\mdot\sim 10^{-4}\,\msunyr$, there may be a significant difference between these 
temperatures. If the wind is so strong that the optical light originates from 
the depth of rapid acceleration, the object is considered to be forming a 
pseudo-photosphere.

The formation of pseudo-photospheres in LBVs may be favored by their lower 
masses, providing an increased mass-loss rate and an 
increase of the photospheric scale-height. Leitherer et al. \cite{Leit89} and de Koter 
et al. \cite{deKoter96} assessed whether variable wind 
properties might explain $\Delta V \simeq 1$ to 2 mags during 
S\,Dor cycles, and concluded 
that pseudo-photospheres are unlikely to form in LBVs. However, they did not 
investigate the effect of an order of magnitude change in the wind density of a 
star that is close to the bi-stability and Eddington limit. 

\begin{figure}
\sidecaption
\includegraphics[scale=.65]{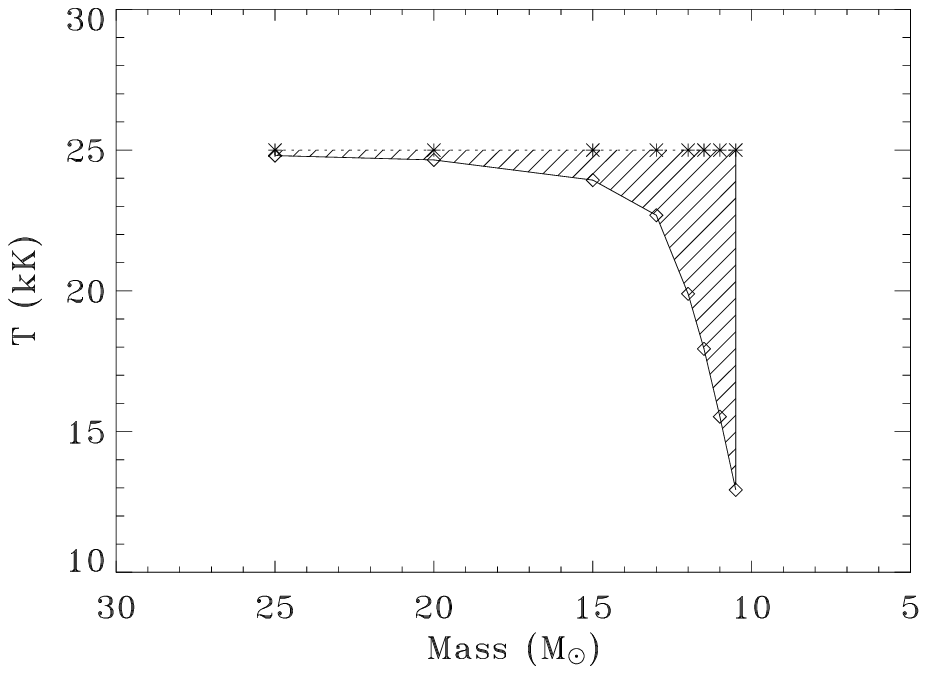}
\caption{The possible formation of a pseudo-photosphere. The figure shows the 
         drop in apparent temperature that results from crossing the bi-stability 
         jump which causes an order of magnitude increase in wind density. The 
         apparent temperature drop from $T_{\rm in} = 25\,000$\,K is shown as a 
         function of the stellar mass -- for objects with log $(L/\lsun)$ $=$ 5.7. 
         The computed effective output temperatures are denoted with $\diamond$.  
         When the stellar mass drops below $M \sim 10.5\,\msun$, the objects starts 
         to form an extended pseudo-photosphere, which results in an effective
         temperature of $\sim$13\,000\,K. Applying a corrective 
         shift of $6\,000$\,K \cite{Vink02} this comes down to  
         $\teff \sim 7\,000$\,K -- corresponding to the location of the YHGs.
         The figure indicates that supergiants that
         have lost significant mass during prior (e.g. RSG) 
         phases may be susceptible to pseudo-photosphere formation. The figure has been adapted from 
         \cite{Smith04}.}
\label{fig:app}
\end{figure}

Smith et al. \cite{Smith04} investigated the 
thermalization optical depth in the inner wind and showed that models on 
the cool side of the bi-stability jump may start to form optically thick winds which 
could lead to the formation of a pseudo-photosphere -- {\it if} the objects are 
close to the Eddington limit.
Figure~\ref{fig:app} shows that for stars with log ($L/\lsun$) $=$ 5.6-5.8, if the current 
mass is below $\sim$12 \msun, corresponding to $\Gamma_{\rm e} = 0.8$, the star would start to form an extended 
optically thick wind when it crosses the bi-stability jump, and for stars  
with  masses below 10.5\,\msun\ (at log ($L/\lsun$) $=$ 5.7) the pseudo-photosphere starts to form at T$_{\rm eff}$ 13000K. 
Applying the $\Delta \teff$ shift of -6\,000 K to match the empirical temperature 
of the bi-stability jump would bring the effective temperature on the cool side 
of the bi-stability jump down to $\sim$7\,000 K, which agrees with the location 
of LBVs in eruption and the YHGs. 

The scenario described above hinges 
critically on the large value of $\Gamma_{\rm e}$ of $>$ 0.8, but it remains 
to be seen if these high $\Gamma_{\rm e}$ values are realistic for LBVs. 
We quoted earlier $\Gamma_{\rm e}$ values in the range $\Gamma_{\rm e}$ $\simeq$ 0.5
for classical, and  $\Gamma_{\rm e}$ $\simeq$ 0.6 for low-luminosity LBVs. It is 
indeed possible that the missing LBVs had larger $\Gamma_{\rm e}$
values, but clearly, more work is needed to determine LBV masses and luminosities 
to reliably establish the proximity of LBVs to the Eddington limit.


\section{Theoretical models for S Dor variability}
\label{sec:theo}

In the previous section we addressed the issue of whether the S~Dor variability 
is the result 
of a sub-photospheric effect that actually increases the radius resulting in a decrease
in T$_{\rm eff}$, or whether the apparent 
decrease in temperature is due to an increase in the mass-loss rate and the
formation of an the optically thick wind. It appears 
that the jury is still out on this, and we should thus consider 
both atmospheric as well as sub-photospheric mechanisms (see also \cite{Hump94}).

In those cases where a pseudo-photosphere 
would form as a result of increased mass loss, the root cause of such a sudden $\dot{M}$ increase 
still needs to be established. 
The bi-stability mechanism is a good candidate because  it can  account for observed LBV 
mass-loss behavior. In the following discussion we include bi-stability
under the more general topic of ``radiation pressure instabilities''.

To investigate which  mechanism might be responsible for the large
visual brightness and spectroscopic variations in S~Dor variables, we must also consider 
the momentum balance defined as:  

    \begin{equation} 
    - \, g_{\mathrm{eff}} \ = \   
       v \frac{dv}{dr}  +  \frac{1}{\rho} \frac{dP}{dr} \; .  
    \end{equation}  
Here $g_{\mathrm{eff}}$ is the net result of gravity, $g_{\mathrm{G}}$, 
minus the outward-directed radiative and turbulent accelerations 
$g_{\mathrm{rad}}$ and $g_{\mathrm{turb}}$.  

\noindent {\bf Radiation pressure instability~~} Luminous evolved stars 
have reduced stability with respect to radiation pressure due to their reduced mass 
for their luminosity and are thus close to the Eddington limit.  However massive stars are also rotating.
The critical velocity is defined as $v_{\mathrm{crit}}^{2} \, = \, (1 - \Gamma) GM/R$.  
When rotation is included via the $v_{\rm rot}^2$ term in the equation of motion, objects may become unstable 
when $\Omega$ $=$ $v_{\rm rot}/v_{\rm crit}$ $>$ 1, 
before arriving at the classical or opacity-modified Eddington limit \cite{Langer97}. 
For example, the projected rotational velocity for  AG\,Car  in the hot phase, 
of 190 $\pm$ 30 \kms\ \cite{Groh06} is close to its critical velocity, and a similar result
was recently found for HR\,Car as well \cite{Groh09b}. 
Thus LBVs are are considered to be in close proximity to both the Eddington 
and the Omega limits. In the sense usually intended, the $\Omega$ limit is really the 
Eddington limit modified by rotation.\footnote
    {In principle this is a simplification of the real physics, 
    because rotation tends to make the equator cooler than the poles 
    (von Zeipel effect).  
    The resulting temperature gradient affects opacity and possible 
    instabilities, potentially making the combination of radiation 
    pressure and rotation synergistic as noted in \cite{LP91,Pel00} as well as section 5 of 
    \cite{Zeth99}} 
Thus, in the following we include the Omega ($\Omega$)
 limit  under the general topic of the ``Eddington limit''.

Radiation-pressure driven instabilities occur because as 
 the temperature drops, the opacity rises (e.g. due to bi-stability), and the
radiative acceleration $g_{\rm rad}$ increases. 
The opacity--modified Eddington limit  was initially introduced 
to explain the great eruption of $\eta$ Car \cite{Dav71}, and subsequently for the temperature dependence of the HD limit \cite{Hump84,Lamers86}, and the instability of LBVs \cite{Ap86,Ap89}.  Lamers \& Fitzpatrick \cite{Lamers88} 
computed the location of                      
the opacity-modified Eddington limit, including metal-line opacities from model
atmospheres in addition to electron scattering. They suggested that S~Dor variations
could result from a conflict between a star's tendency to expand (following
core H-burning) and strong mass loss close to the Eddington limit, requiring
the star to shrink as the mass decreases. However there are issues with this simplified
approach. When the ratio of radiative to gravitational force approaches unity, an instability 
could be expected, but as the 
atmosphere expands and density decreases, the ionization increases thereby reducing the 
absorptive opacity, and instead it approaches the classical Eddington limit due to electron 
scattering (which is not temperature dependent). However, if the instability would occur at 
$\Gamma$ somewhat less than unity, the density decrease would not eliminate absorption, and the concept 
of the modified Eddington limit might nonetheless work.  

The attraction of scenarios based on the Eddington limit are clear; they 
naturally explain  the temperature dependent luminosity limit in the HRD,
the S~Dor variability,  and the two states of LBVs, their high $\teff$ (just on 
the hot side of the bi-stability jump), and their low $\teff$ limits. 
They could also lead to enhanced mass loss, increased density in the winds and
the formation of a pseudo-photosphere. 
However, there are no self-consistent models that provide a sound theoretical basis for 
for scenarios involving pseudo-photospheres \cite{deKoter96}. \\ 

\noindent {\bf Turbulent pressure instability~~}  
As  a star approaches the Eddington limit, the outermost layers of the envelope become convective 
(e.g. \cite{Cant09}) and turbulent pressure gradients may provide an 
additional acceleration, $g_{\rm turb}$ to the momentum equation. 
De Jager \cite{deJager84}
showed how supersonic turbulence may destabilize the atmosphere, and as 
the 
mechanism becomes more efficient at higher luminosity, the mass-loss rate increases.
This is also true for radiation pressure forces and it may  be difficult to 
distinguish
between these two atmospheric instabilities.\\

\noindent {\bf Vibrations and dynamical instability~~} \\ 
Together with the radiation pressure-based instabilities,
sub-photospheric dynamical mechanisms are the most promising explanation
for the LBV/S~Dor variability (see Gr\"afener et al. 2011b and references therein).   
In these models, ``strange modes'' and dynamical instabilities are caused by the
bump or increase in the opacity due to iron at the base of the photosphere leading
to a strong ionization-induced instability in the outer envelope  
as stars transit the HRD after the end of core H-burning.
In the models of Stothers \& Chin \cite{Stothers93} the star 
keeps re-adjusting itself on thermal timescales after periods of strong mass loss, 
whilst shrinking in radius. These models provide the correct S\,Dor timescales 
and also appear to ``behave'' properly 
at constant bolometric luminosity.
The strange mode calculations by Glatzel and Kiriakidis \cite{Glatzel93}  
reproduce the the S~Dor instability strip and the upper luminosity boundary in the 
HRD quite well. Dynamical instability thus remains one
of the more promising candidates to explain LBV variability. 

Vibrational or pulsational instability   
was once thought to be one of the main contenders for instability and mass loss 
in the most massive stars (e.g. \cite{Ap70}), but the $\epsilon$-mechanism   
is energized in the core,  appears to grow too slowly 
(e.g. \cite{Pap73}) and is therefore no longer considered valid for LBVs.
Another sub-photospheric instability, the  $\kappa$ mechanism   
responsible for pulsation in massive stars such as the Cepheids  
 may also cause pulsations in the outer envelope.  It may be  responsible 
for some of the micro-variability seen in LBVs and other supergiants (\cite{Lef07} and references therein) 
which occur on timescales of weeks to months. 
The timescale of the S~Dor variations however is much longer and therefore unlikely 
to be due to pulsations.\\ 

\noindent {\bf Binarity~~}  Most LBVs are apparently single, and although.
 $\eta$\,Car may have a companion, it seems clear that binarity can neither be 
the root cause of the S\,Dor variations, or for the giant eruptions, as the only 
other local example, 
P\,Cyg, is single (but see \cite{Kashi10}, or unless it formed through 
merging, cf. \cite{Podsi90}).


\section{Evolutionary State}
\label{sec:sn}

There is no doubt that LBVs are evolved, unstable massive hot stars. 
The more massive classical LBVs have apparently evolved off the main sequence, while the less luminous LBVs may be 
post-red supergiants. 
In the generally accepted view of massive star evolution,
the classical LBVs are considered ``transitional'' objects in a phase before
entering the He-burning WR stage \cite{Langer94}, by the end of which
the star is anticipated to explode as a type Ib/c supernova.
Many  LBVs are known to be N and He rich compared to O stars, but H rich compared to
the more evolved WR stars. This situation is somewhat more complex
as there is also a group of high-luminosity late-type H-rich WR stars,
which appear closely related to many classical LBVs in quiescence the Ofpe/WN stars 
\cite{Walborn82}.  These and other luminous stars, the B[e] supergiants \cite{Lamers98}, and
and the cool or yellow Hypergiants (YHGs) \cite{deJager98}, that show evidence for 
high mass loss and instabilities which may be related to the LBV state.

\subsection{The evolutionary neighbours}

\noindent {\bf Ofpe/late-WN stars~~} These ``slash'' stars are a group of luminous hot 
stars with very strong emission lines due to their  strong mass 
loss. They  
have He and N enhanced atmospheres indicative of  an evolved 
state  \cite{Pasquali97}, but are  generally not believed 
to be highly variable. These stars however may be closely related to the LBVs in quiescence.
The S~Dor-type variable, R127, was a late-WN star prior to its long-term outburst 
 beginning in the early 1980s \cite{Stahl83}. Either 
the late-WN stars  evolve {\it into} LBVs, or they may represent a dormant phase of 
LBV evolution. Either way, Ofpe/late-WN stars are thought to 
be evolved massive stars   in a transitional stage for objects with initial masses 
$M$ $>$ 50-60 $\msun$.  Since  no evolved stars  are observed  redwards of the Humphreys-Davidson (HD) 
limit at these high luminosities and masses , high  mass loss during the Ofpe/late-WN and LBV phases may reverse 
the evolutionary track  back to the hotter part of the HRD, where they should appear 
as He-burning WR stars.\\

\noindent {\bf B[e] supergiants~~} The spectra of the B[e] stars \cite{Lamers98} 
show an abundance of  high-excitation permitted  and forbidden emission 
lines that are thought to arise from an equatorially enhanced outflowing disk. 
Zickgraf et al. \cite{Zickgraf86} proposed a 2-component wind with a normal fast polar wind and 
a dense slow outflowing equatorial ``disk''. A popular mechanism to explain this 
2-component wind is the rotationally induced bi-stability mechanism  
 \cite{LP91,Pel00}. The pole 
is hotter than the equator, due to the Von Zeipel gravity darkening effect which 
could lead to a fast, low $\dot{M}$, polar wind driven by Fe {\sc iv} lines, and a 
slow, high $\dot{M}$, equatorial wind driven by the more effective Fe {\sc iii} 
lines \cite{Vink99}. This mechanism is expected to  
occur predominantly at spectral type B. However  the 
star is expected to rotate rapidly but $v_{\rm rot}$ measurements  are difficult for B[e] 
stars, with most of the lines in emission. 

The B[e] 
supergiants may represent a subset of massive stars with high rotational 
velocities. The B[e] supergiants were originally not thought to be variable,  but  
there is now  evidence for   large 
amplitude variability for some B[e] supergiants \cite{vanG02}, 
 suggesting  a closer evolutionary link between LBVs and B[e] supergiants 
than previously acknowledged.\\

\noindent {\bf Yellow or cool hypergiants~~} YHGs are found just below the HD-limit \cite{Hump79,Hump94} at intermediate  temperatures with A to G spectral types.  Many of these stars  
show spectroscopic and photometric variability, high mass-loss rates,  large infrared excesses and visible circumstellar ejecta, all evidence for instability.  
The YHGs are often assumed to be post-RSGs \cite{deJager98,Oud08}, although  the 
evolutionary state is not established for all of them.  
The intriguing object IRC $+$10420 has been shown to be a post-RSG \cite{Jones93,Oud96} 
and numerous
studies have revealed its complex ejecta \cite{Hump97,Hump02} and large-scale asymmetry 
 \cite{Patel08,deWit08,Davies07b}. Given their variability and high mass loss, the YHGs
are likely close to their Eddington limit, with a large $\Gamma_{\it e}$, and it
is thus  probable that many of them  are post-RSGs 
Nevertheless,  it is not clear whether  objects like IRC $+$10420 are on an 
evolutionary blueward journey towards the WR phase \cite{Oud96,Hump02}  
 bouncing against the yellow void \cite{deJager98}, or are on the cool side of the 
bi-stability jump \cite{Smith04}.   

\subsection{Do LBVs explode?}

The picture of the LBVs as a high mass loss, relatively short-lived (some $10^{4}$ yrs) 
and presumably core H-burning phase prior to a much longer (a few 
times $10^{5}$ yrs)
core He-burning WR phase seemed well established -- until recently.

There is increasing observational evidence that LBVs 
could be {\it direct} progenitors of some SNe. 
Kotak \& Vink \cite{Kotak06} proposed that the quasi-periodic
modulations seen in the radio light curves of transitional SNe such as SN 2001ig and SN 2003bg 
are the manifestation of variable mass loss during S\,Dor excursions. 
Although several other possibilities have been put forward to explain these 
modulations \cite{Ryder04,Soder06}, none has 
been entirely satisfactory. 
The recurrence timescale
of the variability, as well as the amplitude of the radio modulations are consistent with  
 those of S\,Dor variables and their scenario \cite{Kotak06} provides a rather natural
explanation for a behaviour that is expected on theoretical grounds \cite{Vink02}. 

\begin{figure}
\sidecaption
\includegraphics[scale=.45,angle=90]{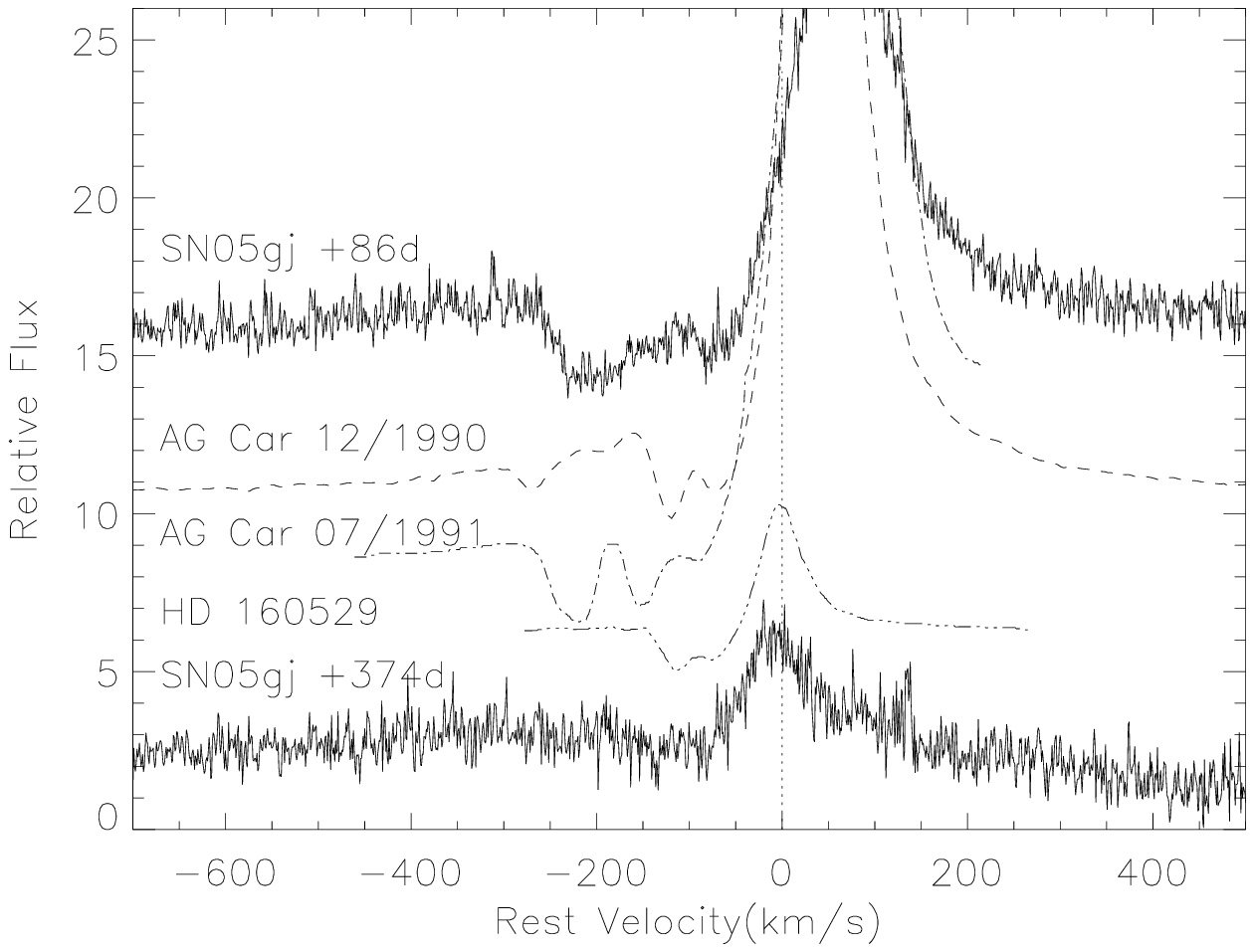}
\caption{Multiple absorption components seen in the P Cygni H$\alpha$ profile 
of SN 2005gj (top) in comparison to the LBVs AG~Car and HD\,160529.
The figure has been taken from Trundle et al. \cite{Trundle08}.} 
\label{fig:trundle}
\end{figure}

The same wind bi-stability mechanism may be 
able to account for wind-velocity variations seen spectroscopically in SN 2005gj 
\cite{Trundle08} in which 
 the variable winds are inferred from double P~Cygni components (see Fig.~\ref{fig:trundle}) 
which appear almost identical to those seen in the H$\alpha$ profiles of S\,Dor variables like 
AG\,Car and HD\,160529 (see also \cite{groh11}). Both the timescales and the 
spectroscopically measured wind velocities of SN 2005gj, with $\vinf$ $\simeq$100-200 \kms, 
are consistent with those of LBVs, but  not  with those 
of the much slower RSG winds ($\sim$10 \kms), or the much faster WR winds ($\simeq$1000-5000 \kms).
See Van Marle et al. \cite{vanM07} and \cite{Trundle08} for more information.

The progenitor star of the recent supernova SN~2006jc had a  
giant eruption just two years before its terminal explosion.
Foley et al. \cite{Foley07} and Pastorello et al. \cite{Past07} suggested that the 
progenitor star was either a WR star that exhibited an ``LBV or $\eta$ Car-like'' eruption, 
or that the progenitor was part of a 
binary system including both an LBV and a WR star, with the WR star exploding and the giant eruption 
attributable to the LBV, as WR stars have never been observed to have an 
$\eta$\,Car type eruptions.
A more direct application of Occam's razor would be to accept that the progenitor 
object exploded during or at the end of the LBV phase or an $\eta$ Car-like giant eruption  \cite{Kotak06}.

Gal-Yam et al. \cite{GalYam07} 
discovered a luminous source (with $M_{\rm bol}$ $=$ $-$10.3) in the 
pre-explosion image of SN~2005gl. 
Although,  the properties of  the progenitor  
are  consistent
with those of LBVs, they are equally consistent with  a  
luminous blue supergiant that has not exhibited spectral type variations 
(see the membership discussion in Sect.~\ref{sec:props}),  and 
 the progenitor of SN~2005gl may potentially be classified  as an LBVc. 
 The resulting SN explosion was of type IIn, indicating the presence of  
a dense circumstellar medium, additional evidence for  
 prior $\eta$\,Car-type eruptions.

Other hints that LBVs may explode come from similarities in the morphology of 
LBV nebula and the circumstellar medium  of SN\,1987A \cite{Smith07}, while the 
very luminous 
SN~2006gy -- hypothesized to be an exotic pair-instability SN -- may also have undergone 
an $\eta$\,Car-type eruption before exploding \cite{Smithea07}.
An alternative scenario for a giant $\eta$ Car-type outburst was 
suggested by Woosley et al. \cite{Woosley07} who attributed the dense shells around 
this luminous SN to pulsational pair instability. 

We emphasize that  the 
evolutionary status of LBVs remains uncertain (see the discussion in \cite{Dwarkadas11}). 
Evolutionary models have been constructed to allow LBVs in a transitional 
phase between a core H-burning main sequence and a core He-burning 
WR phase. This model   
naturally accounts for the 
the chemical abundances (He, N) of LBVs which are 
intermediate between those of O and WR stars. 
The concept of an LBV exploding is certainly at odds with current 
stellar evolution models and an exploding LBV 
scenario was until recently considered ``wildly speculative'' \cite{Smith06}.
Nevertheless, the evolutionary models do not provide a straightforward explanation for  
the wide range of phenomena described above and a simple explanation for this 
could be that at least  some massive stars in  an LBV state  could precede the terminal
explosion.  This would allow some massive stars  to skip the WR phase -- in contradiction 
to the basic framework of massive star evolution.  
This also suggests  that LBVs are already in the core He-burning phase of evolution. 
Given the intriguing variations seen in radio lightcurves and especially the  
double absorptions seen spectroscopically in P\,Cygni line profiles, I suggest that 
the changing winds of LBVs may help addressing the issue of whether the LBV phase may 
indeed represent the evolutionary endpoint for some of the most massive stars.


\section{Outlook}
\label{sec:outlook}

We have examined observational, atmospheric modelling, and 
theoretical aspects of the current status of our knowledge of the LBV instability
and their role in massive star evolution. 
Although radiation pressure as well as a dynamical instability  are 
strong candidates for explaining S\,Dor variability, we 
conclude that the mechanism at the origin of the LBV phenomenon
remains elusive. One of the most relevant issues to 
be addressed relates to the nature of the S\,Dor variability itself. 
With respect to massive star evolution, and in particular whether LBVs are in a transitional or final phase 
of evolution, the last few years have seen a flurry of activity, but we should not yet draw 
any definitive conclusions regarding their evolutionary state.

Progress is expected to be made on a number of fronts. First of all, the number 
of known LBVs is small and for a proper understanding of any individual member, 
photometric, polarimetric, and spectroscopic monitoring on the timescale of the 
S\,Dor variations is required. Atmospheric modelling is necessary to determine 
the stellar parameters and place the LBVs properly onto the HRD.  
There are still enormous uncertainties in the values shown in Fig.~\ref{fig:hrd}.
Furthermore, LBV masses, nor their proximity to the Eddington 
limit, are known with any level of certainty, with profound consequences for 
theoretical interpretation. Therefore, binary searches, orbit 
determinations, and spectroscopic modelling are strongly encouraged 
to determine this most basic parameter. 

As is the case for the normal OB supergiants, the LBV mass-loss rates are 
uncertain due to wind clumping. Progress will undoubtedly be made regarding the role of 
of wind clumping and its impact on mass-loss rates in OB stars and LBVs.  
However, the LBVs themselves provide an ideal 
laboratory for studies of wind clumping, because the polarization variability is 
most extreme, due to the combined effect of low outflow velocities and high 
mass-loss rates \cite{Davies05}. Mass-loss variability -- in conjunction with 
$\teff$ determinations -- could also be utilized to constrain the stellar masses, as was 
exemplified for the object AG\,Car \cite{Vink02}.

A number of developments provide new opportunities for 
LBV research. The infrared era has properly come of age with satellites such as {\it Spitzer}. 
Recent activities involve both in-depth studies of known LBVs \cite{Umana09}, as well as the 
discoveries of new ones, even 
beyond the local group. We highlight the so-called intermediate luminosity transients, 
where ``intermediate'' means that their luminosities are in between 
classical novae and supernovae. The nature of these transients, incl. SN 2008S and NGC 300 OT2008-1, 
remains as yet elusive. Some authors argue for a SN origin \cite{Prieto08,Bot09}, whilst 
others \cite{Berger09,Bond09,Smith09} suggest they are due to the non-terminal outburst of an 
LBV-type star.
Another intriguing direction for LBV research is provided by the detection of 
LBVs and candidates in very low metallicity environments \cite{Pus08,Izotov09,Herrero10}. The mere 
presence of LBVs at such low metallicities presents a challenge for 
theory, because of the reduced opacity available to drive an LBV wind.  

Further theoretical work on the formation of pseudo-photospheres and the more 
general question regarding the origin of the LBV variations are badly needed. 
Such studies may result in a better understanding of the origin of the variability 
which is necessary to place the LBV phase correctly in the evolution 
of massive stars. 

The final message to emerge from this chapter is that $\eta$ 
Car may be one of the most extreme LBVs but it is not 
unique among the LBVs. 
Given that $\eta$ Car's second outburst (during 1888-1895) 
was like that of a normal S\,Dor variable, the key to our understanding 
of $\eta$\,Car's great eruption may thus not exclusively lie in the study of 
$\eta$\,Car itself, but pivotal clues may be obtained through a better 
understanding of the more typical S\,Dor variations 
that define the LBVs as a class among the most massive stars.

{\it The most outstanding question for all these stars, S~Dor-type LBVs and the
giant eruptions like $\eta$ Car, is still, what is the underlying origin of their
instabilities?}

\begin{acknowledgement}
I am grateful for the many discussions I have had with colleagues from the massive star
community and in particular to Alex de Koter, Henny Lamers, and Rubina Kotak.
\end{acknowledgement}

\end{document}